\newcommand{\beq}{\begin{equation}}
\newcommand{\eeq}{\end{equation}}

\newcommand{\id}
 {i\kern.06em\hbox{\raise.25ex\hbox{$/$}\kern-.60em$\partial$}}

\newcommand{\bs}{/\kern-.52em b}
\newcommand{\qs}{/\kern-.52em s}

\newcommand{\yp}{^{\prime}}

\newcommand{\dd}
{\kern.06em\hbox{\raise.25ex\hbox{$/$}\kern-.60em$\partial$}}

\documentstyle[12pt]{article}
\date{}
\begin{document}
\title{A Note on the Proof of Magnetic Flux Quantization 
from ODLRO
\footnotetext{\# Corresponding adress}
\thanks{On leave of absence from the Physics Department of Shanghai
University, 201800, Shanghai, China}}
\author{{Feng Sze-Shiang $^{1,2}$, Zhu Dong-Pei $^2$}\\
1. {\small {\it High Energy Section, ICTP, Trieste, 34100, Italy}}\\
e-mail:fengss@ictp.trieste.it\\
1.{\small {\it CCAST(World Lab.), P.O. Box 8730, Beijing 100080}}\\
2.{\small {\it Department of Modern Physics , University of Science
and Technology of China, 230026, Hefei, China}}$^\#$\\e-mail:zhdp@
ustc.edu.cn$^\#$}
\maketitle
\newfont{\Bbb}{msbm10 scaled\magstephalf}
\newfont{\frak}{eufm10 scaled\magstephalf}
\newfont{\sfr}{eufm7 scaled\magstephalf}
\input amssym.def
\baselineskip 0.3in
\begin{center}
\begin{minipage}{135mm}
\vskip 0.3in
\baselineskip 0.3in
\begin{center}{\bf Abstract}\end{center}
  {It is noticed that the excellent proof of the connection
  of magnetic flux quantization and off-diagonal long range
  order(ODLRO) presented recently by Nieh, Su and Zhao suffers
  from an imperfection, namely, the f-factors in the case of finite
  translation do not satisfy $f({\bf a})f({\bf b})=f({\bf a+b})$,
  which was imployed in the proof. A corrected proof
  is proposed to remedy this  point.
  \\PACS number(s): 75.10.Lp,71.20.Ad,74.65.+n,74.20.Mn
   \\Key words: flux quantization, ODLRO}
\end{minipage}
\end{center}
\vskip 1in
The study of ODLRO of fermions, especially those strongly correlated,
has drawn much attention recently\cite{s1}
-\cite{s16} since the onset of ODLRO implies a phase transition in the system.
This trend of investigations is currently believed to be able to
sheed some light on the explanation of the high-$T_c$
superconductivity in cuprates by a number of physicists.
In his original paper\cite{s17},
Yang argued that the existence of ODLRO
will likely lead to Meisner effect and magnetic flux quantization. The key
spirit of the argument asserts that the dependence of the thermodynamic
partition function on the flux, which provides  phase factors responsible
for space tranlations of the reduced density matrices , is very similar
to the dependence of Bloch wave function
on the wave vector in crystals. Although this argument is intuitive, it is
not clearcut and direct. Quite recently, Sewell\cite{s18} gave a proof
of the relationship of Meisner effect and ODLRO. Later, Nieh, Su and Zhao
\cite{s19} extended the proof to the case of flux quantization. Yet,
their extention suffers from an imperfection when the lattice potential
is taken into consideration. This note is to remedy this point.\\
\indent In general, the reduced density matrix $\rho_2({\bf r}_1\yp, {\bf r}
_2\yp;{\bf r}_1,{\bf r}_2)$ can enjoy a spectral seperation\cite{s17}
\beq
\rho_2({\bf r}_1\yp, {\bf r}_2\yp;{\bf r}_1,{\bf r}_2)=\sum_s\alpha_s\Phi_s
({\bf r}_1\yp, {\bf r}_2\yp)\Phi_s^*({\bf r}_1,{\bf r}_2)
\eeq
where $\Phi_s({\bf r}_1,{\bf r}_2)$ are normalized eigenfunctions
of $\rho_2$ with eigenvalues $\alpha_s$:
\beq
\int\Phi^*_s({\bf r}_1,{\bf r}_2)\Phi_s({\bf r}_1,{\bf r}_2)
d{\bf r}_1d{\bf r}_2=1
\eeq
\beq
\int\rho_2({\bf r}_1\yp, {\bf r}_2\yp;{\bf r}_1,{\bf r}_2)d{\bf r}_1d{\bf r}
_2\Phi_s({\bf r}_1,{\bf r}_2)=\alpha_s\Phi_s({\bf r}_1\yp,{\bf r}_2\yp)
\eeq
The existence of ODLRO in $\rho_2$ means that in the off-diagonal long
range (ODLR) limit, i.e., $\mid{\bf r}_i\yp-{\bf r}_j\mid\rightarrow\infty$
while keeping $\mid{\bf r}_1-{\bf r}_2\mid$ and $\mid{\bf r}_1\yp-
{\bf r}_2\yp \mid$ finite, $\rho_2$ behaves as
\beq
\rho_2({\bf r}_1\yp, {\bf r}_2\yp;{\bf r}_1,{\bf r}_2)\rightarrow
\alpha\Phi({\bf r}_1\yp, {\bf r}_2\yp)\Phi^*({\bf r}_1,{\bf r}_2)
\eeq
where $\alpha$ is the largest eigenvalue which is of order $O(N)$. The
important observation made in\cite{s19} is that the space translation
${\bf r}\rightarrow{\bf r-a}$ will effect a gauge transformation, based on which
two complete sets of basis functions were obtained. As a consequence, the
transform property of $\rho_2$ is
\beq
\rho_2({\bf r}_1\yp, {\bf r}_2\yp;{\bf r}_1,{\bf r}_2)=\exp\{
\frac{ie}{\hbar c}[\chi_{{\bf a}}({\bf r}_1\yp)+\chi_{{\bf a}}({\bf r}_2\yp)
-\chi_{{\bf a}}({\bf r}_1)-\chi_{{\bf a}}({\bf r}_2)]\}
\rho_2({\bf r}_1\yp-{\bf a}, {\bf r}_2\yp-{\bf a};{\bf r}_1-{\bf a}
,{\bf r}_2-{\bf a})
\eeq
where $\chi_{{\bf a}}({\bf r})={\bf a}\cdot{\bf A}_0({\bf r})
+\phi({\bf r}-{\bf a})-\phi({\bf r}), {\bf A}_0=\frac{1}{2}{\bf B}\times
{\bf r}, \phi({\bf r})$ is a gauge function which is in general multivalued.
Therefore, in the ODLR limit, we have
\beq
\Phi({\bf r}_1\yp, {\bf r}_2\yp)\Phi^*({\bf r}_1,{\bf r}_2)
\rightarrow\exp\{
\frac{ie}{\hbar c}[\chi_{{\bf a}}({\bf r}_1\yp)+\chi_{{\bf a}}({\bf r}_2\yp)
-\chi_{{\bf a}}({\bf r}_1)-\chi_{{\bf a}}({\bf r}_2)]\}
\Phi({\bf r}_1\yp-{\bf a}, {\bf r}_2\yp-{\bf a})\Phi^*({\bf r}_1-{\bf a}
,{\bf r}_2-{\bf a})
\eeq
which implies
\beq
\Phi({\bf r}_1, {\bf r}_2)=f({\bf a})
\exp\{\frac{ie}{\hbar c}[\chi_{{\bf a}}({\bf r}_1)+\chi_{{\bf a}}({\bf r}_2)
]\}\Phi({\bf r}_1-{\bf a}, {\bf r}_2-{\bf a})
\eeq
where $f){\bf a})$ is a position-independent phase factor, $f(0)=1$. So
$\Phi$ differ by a phase factor at different space points.\\
\indent Let us consider any two {\it finite} successive space translations
, firstly by ${\bf a}$ and then by ${\bf b}$. Since
\beq
\chi_{{\bf b}}({\bf r}-{\bf a})={\bf b}\cdot{\bf A}_0({\bf r}-{\bf a})
+\phi({\bf r}-{\bf a}-{\bf b})-\phi({\bf r}-{\bf a})
\eeq
we have
\beq
\chi_{{\bf a}}({\bf r})+\chi_{{\bf b}}({\bf r}-{\bf a})=({\bf a}+{\bf b})
\cdot {\bf A}_0({\bf r})+\phi({\bf r}-{\bf a}-{\bf b})-\phi({\bf r})-
\frac{1}{2}{\bf b}\cdot{\bf B}\times {\bf a}=\chi_{{\bf a}+{\bf b}}({\bf r})
-\frac{1}{2}{\bf b}\cdot{\bf B}\times {\bf a}
\eeq
Accordingly,
\beq
\Phi({\bf r}_1, {\bf r}_2)=f({\bf a})f({\bf b})
\exp\{\frac{ie}{\hbar c}[\chi_{{\bf a}+{\bf b}}({\bf r}_1)
+\chi_{{\bf a}+{\bf b}}({\bf r}_2)]\}
\exp\{[\frac{ie}{\hbar c}{\bf b}\cdot{\bf a}\times{\bf B}]\}
\Phi({\bf r}_1-{\bf a}-{\bf b}, {\bf r}_2-{\bf a}-{\bf b})
\eeq
Thus for a closed curve $C$, which is in general zigzag when the translation
vectors ${\bf a}, {\bf b}, ...$ are finite, ${\bf a}+{\bf b}+...=0$, eq.(10)
together with the singlevaluedness of $\Phi$, which is ensured by the single-
valuedness of $\rho_2$, give
\beq
f({\bf a})f({\bf b})...\exp\{\frac{2ie}{\hbar c}{\bf B}\cdot{\bf S}\}=1
\eeq
where ${\bf S}$ is the area enclosed by $C$ with the orientation of the
translation vectors. It is obvious that
\beq
f({\bf a})f(-{\bf a})=1
\eeq
Now suppose that the three primitive vectors of the crystal  are
${\bf a}_i, i=1,2,3.$ Consider now that the closed path constituted by
${\bf a}_i, {\bf a}_j. -{\bf a}_i$ and $-{\bf a}_j$, we have then
\beq
f({\bf a}_i)f({\bf a}_j)f(-{\bf a}_i)f(-{\bf a}_j)=\exp\{\frac{-2ie}{\hbar c}
{\bf B}\cdot{\bf S}\}
\eeq
The quantization of flux follows then from (12)
\beq
{\bf B}\cdot{\bf S}=n\frac{2\pi\hbar c}{2e}
\eeq
Since the size of such  a region enclosed by any two primitive vetors is of the
typical order $(1\AA)^2$, if $n\not=0$, $B$ must be of the order of
magnitude $B\sim(1\AA)^{-2}\frac{2\pi\hbar c}{2e}\sim 10^9$ G (Note
that ${\bf B}$ is usually parallel to ${\bf S}$ in experiments), which is much stronger than
the typical critical field strength of less than $10^3$ G, therefore
$n$ must be zero, hence ${\bf B}=0$. This states that if ${\bf S}$ is
simply connected, i.e., the region enclosed by $C$ is superconducting everywhere,
we have Meisner effect. On the other hand, if ${\bf S}$ is multiconnected,
i.e., it contains normal regions which are of mesoscopic sizes, $n$ may be
non-zero integers. In this case, we have the quantization of magnetic flux.\\
\indent We can also draw the same conclusion in another way. Since (10) can be
also expressed as
\beq
\Phi({\bf r}_1, {\bf r}_2)=f({\bf a}+{\bf b})
\exp\{\frac{ie}{\hbar c}[\chi_{{\bf a}+{\bf b}}({\bf r}_1)
+\chi_{{\bf a}+{\bf b}}({\bf r}_2)]\}
\Phi({\bf r}_1-{\bf a}-{\bf b}, {\bf r}_2-{\bf a}-{\bf b})
\eeq
we have
\beq
f({\bf a}+{\bf b})=\exp\{[\frac{ie}{\hbar c}{\bf B}\cdot({\bf b}\times{\bf a})
]\}f({\bf a})f({\bf b})
\eeq
and similarly
\beq
f({\bf b}+{\bf a})=\exp\{[\frac{ie}{\hbar c}{\bf B}\cdot({\bf a}\times{\bf b})
]\}f({\bf b})f({\bf a})
\eeq
Since $f({\bf a}+{\bf b})=f({\bf b}+{\bf a})$ and $f({\bf a})$ are c-numbers,
we infer that
\beq
\exp\{\frac{2ie}{\hbar c}{\bf B}\cdot({\bf a}\times{\bf b})=1
\eeq
Therefore, for any two lattice vectors ${\bf a}$ and ${\bf b}$, we always
have
\beq
{\bf B}\cdot({\bf a}\times{\bf b})=n\frac{2\pi\hbar c}{2e}
\eeq
\indent The imperfection of the proof in\cite{s19} is that, when 
the lattice potentials are taken into account, the translation vectors can only
be lattice vectors. But for lattice vectors ${\bf a}$ and ${\bf b}$,
the multiplication rule is (16) instead of the following
\beq
f({\bf a}+{\bf b})=f({\bf a})f({\bf b})
\eeq
which was used in \cite{s19} as the basis to prove the flux quantization. Eq(20) holds
only for infinitesimal translations if neglecting second order infinitesimals.
\\
\indent The relation is somewhat similar to that of magnetic translation
operators which realize a ray representation of the translation group
\cite{s20}. Consider the Hamiltonian of $N$ electrons
\beq
H=\sum^N_{i=1}\frac{1}{2m}[{\bf P}_i+\frac{e}{c}{\bf A}_0({\bf r}_i)]^2
+\frac{1}{2}\sum_{i,j}V({\bf r}_{ij})+\sum_iV_c({\bf r}_i)+\sum
_i\frac{e\hbar}{mc}{\bf s}_i\cdot{\bf B}
\eeq
Define ${\bf \beta}=\frac{e{\bf B}}{\hbar c}$ . For any lattice vector
${\bf R}_m=m_1{\bf a}_1+m_2{\bf a}_2+m_3{\bf a}_3$, consider the
operator
\beq
{\cal T}({\bf R}_m)=\exp\{\sum^N_{j=1}\frac{i}{\hbar}{\bf R}_m\cdot{\bf P}
_j\}\exp\{i\sum^N_{j=1}({\bf \beta}\times{\bf R}_m)\cdot\frac{{\bf r}_j}{2}\}
\eeq
We have ${\cal T}{\cal T}^\dag={\Bbb I} $ and$ [{\cal T}({\bf R}_m), H]=0$. Since
\beq
{\cal T}({\bf R}_m){\cal T}({\bf R}_n)=\exp\{i\frac{1}{2}N{\bf \beta}
\cdot({\bf R}_n\times{\bf R}_m)\}{\cal T}({\bf R}_m+{\bf R}_n)
\eeq
we have
\beq
{\cal T}({\bf R}_m){\cal T}({\bf R}_n)=\exp\{iN{\bf \beta}
\cdot({\bf R}_n\times{\bf R}_m)\}{\cal T}({\bf R}_n){\cal T}({\bf R}_m)
\eeq
This is what (16) is similar to. As the Bloch wave functions can be
designated by the eigenvalues of the translation operators, i.e., the
wave vectors ${\bf k}$ and the Hamiltonian, in the presence of magnetic
field, the wave functions can be characterized by the eigenvalues of the
Hamiltonian and the magnetic translation operators belonging to any
chosen commutable set ${\cal R}=\{{\bf R}\mid[{\cal T}({\bf R}_m),
{\cal T}({\bf R}_n)]=0\}.$
\vskip 0.3in
\underline{\bf Acknowledgement} S.S. Feng is grateful to Prof. S. Randjbar-Daemi for
his invitation to ICTP for three months. This work was
supported by the Funds for
Young Teachers of Shanghai Education Commitee and in part by the National
Science Foundation of China under Grant No. 19805004 and No. 19775044.\\

\end{document}